\documentclass[useAMS,usenatbib]{mn2e}
\usepackage{lscape}
\usepackage{graphicx}

\title[A Catalogue of RR Lyrae Stars from the NSVS]{A Catalogue of RR Lyrae Stars from the Northern Sky Variability Survey}
\author[P. Wils, C. Lloyd \& K. Bernhard]{Patrick Wils$^{1}$, Christopher Lloyd$^{2}$, Klaus Bernhard$^{3,4}$\\
$^{1}$Vereniging voor Sterrenkunde, Belgium, email: patrick.wils@cronos.be \\
$^{2}$Space Science \& Technology Department, Rutherford Appleton
Laboratory, Chilton, Didcot, Oxon. OX11 0QX, UK, email: cl@astro1.bnsc.rl.ac.uk \\
$^{3}$A-4030 Linz, Austria, email: klaus.bernhard@liwest.at \\
$^{4}$Bundesdeutsche Arbeitsgemeinschaft f\"ur Ver\"anderliche Sterne
e.V. (BAV), Munsterdamm 90, D-12169 Berlin, Germany}

\begin{document}

\date{Accepted 2006 February 23. Received 2006 February 21; in original form 2006 January 16}

\pagerange{\pageref{firstpage}--\pageref{lastpage}} \pubyear{2006}

\maketitle

\label{firstpage}

\begin{abstract}
A search for RR Lyrae stars has been conducted in the publicly available
data of the Northern Sky Variability Survey ({\it NSVS}). Candidates have been
selected by the statistical properties of their variation; the standard
deviation, skewness and kurtosis with appropriate limits determined from a
sample 314 known RRab and RRc stars listed in the GCVS.  From the period analysis and light curve shape of over 3000
candidates 785 RR Lyrae have been identified of which 188 are previously
unknown. The light curves were examined for the Blazhko effect and several
new stars showing this were found. Six double-mode RR Lyrae stars were
also found of which two are new discoveries. Some previously known
variables have been reclassified as RR Lyrae stars and similarly some
RR Lyrae stars have been found to be other types of variable, or not
variable at all.
\end{abstract}


\section{Introduction}

Statistical studies of the relative numbers of the different classes of RR Lyrae variable stars
and especially the incidence rate of multi-periodicity may give indications
of the metallicity of different stellar systems and of their evolution \citep[see e.g. ][]{mos}.
Exhaustive studies have been done in the Magellanic Clouds by the {\it MACHO} collaboration \citep[][ and 2003]{macho}
and the {\it OGLE} survey \citep{oglelmc}.
Other studies have searched for RR Lyrae stars in the Galaxy, 
such as {\it QUEST} \citep{quest} and also {\it OGLE} \citep{collinge}.  
Most of the stars found in these studies are faint, 
and limited to a small region of sky (the galactic bulge and the equator for {\it OGLE} and
{\it QUEST} respectively).  
The Robotic Optical Transient Search Experiment \citep[{\it ROTSE-1}, ][]{rotse} 
found a fairly large number of previously unknown bright 
(magnitude $< 15$) RR stars in a part of the sky.

This paper sets out to extend this search for field galactic RR Lyrae stars 
to the whole northern sky in the {\it ROTSE-1} data, 
made publicly available via the Internet \citep[Northern Sky Variability Survey - {\it NSVS},][]{skydot}.

\section[]{Methodology}

With only photometric data, and no spectral information, the type of a short period variable can only be determined once the (phased) light curve is known,
and hence only once the period is known.  However, determining the period from sparse data is
a computationally demanding process.  Therefore it was decided to limit
the number of objects by statistical parameters involving much less computation: standard deviation,
skewness, kurtosis and the mean square of successive differences of the magnitude data.

By looking at the values for these statistics for the RR Lyrae stars listed in the 
General Catalogue of Variable Stars \citep[GCVS, ][ and its online edition]{gcvs},
limiting conditions were then derived.  The aim was to set these limits as strict as
possible, so that not too many objects needed to be checked, but still to include the majority of
RR Lyrae stars.
Those GCVS stars that did not follow the criteria, can then provide an estimate for the
completeness of the survey.

\subsection[]{Data}

The {\it ROTSE-1} was an unfiltered CCD survey of the sky from the north pole
to declination $\sim -38$, reaching
magnitude $\sim 15$ with varying levels of completeness. The survey lasted
nominally for one year but depending on the circumstances coverage of
individual objects may be significantly less than this. Objects typically
have 100 to 500 measurements with a median photometric accuracy of 0.02 mag for 10th magnitude stars
and a positional accuracy of 2". The spatial resolution of 14"
compromises the photometry in crowded fields, typically with $|b| < 20$, but
also at higher galactic latitudes for stars with companions within $\sim
45"$.

The data are publicly available from the Sky Database for Objects in
Time-Domain (SkyDOT) web site \citep{skydot} and it is possible to select data with
respect to 8 extraction flags and 7 photometric correction flags. The
default selection for good data sets all but one, {\sc PATCH}, of the
photometric correction flags and only one, {\sc SATURATED}, of the extraction
flags. However, experience of working with the data has shown that
observations with the extraction flag {\sc APINCOMPL} set are often completely
out of range and should be rejected. On the other hand data with the
photometric correction flag {\sc RADECFLIP} set are often indistinguishable from
the other data. So the data have been selected with the {\sc SATURATED} and
{\sc APINCOMPL} flags set and the {\sc PATCH} and {\sc RADECFLIP} flags unset.

It was also decided that 
only stars with 100 good {\it NSVS} observations or more were to be considered,
in order to get good statistics and reliable period determinations, 
and to possibly detect multiperiodicity (double-mode pulsation or a Blazhko effect).
With fewer data points, statistics may be influenced substantially by erroneous observations,
such as those introduced by e.g. a close companion.
Also, it is then not always possible to derive the correct period:
in view of the sampling frequency, 
and the rather short total time span of the available data (less than a year) alias frequencies will be more important.

RR Lyrae stars are fairly blue stars (spectral types A and F).  
The {\it NSVS} survey however observed only in one colour (unfiltered CCD), so colour information has to be retrieved
from another source, such as {\it 2MASS} \citep{2mass}.  
The {\it NSVS} positions are not very accurate however, and
matching them to {\it 2MASS} coordinates may be troublesome in crowded fields.  
In view of this and of reddening aspects, it was decided not to use colour information as a filter.

\subsection[]{Control group: GCVS stars}

The GCVS stars that were to be considered for the control group, had to be well-known and have an accurate position.
Therefore only the GCVS types RRab or RRc were taken (no RR, RR:, RRab: or RRc: stars, i.e. the GCVS classification 
should be precise enough to give the exact subtype).  
Because of the limited number of RRd stars in the GCVS (the RR(B) class), 
these were not taken into account either.
For practical purposes, as far as their statistical parameters are concerned, these double-mode stars can be 
considered to be RRc stars.
The known RR Lyrae stars in the GCVS were further limited to the constellations And to Ori 
for which precise positions had been determined by the GCVS team at the time this study started.  
Many stars in other constellations did not have accurate enough coordinates, 
which could lead to misidentifications with the {\it NSVS} stars.  
It is however still possible that a faint RR Lyrae star unobservable 
by the {\it ROTSE} camera, lies close to a brighter (constant) companion, leading to a false identification.
Because of this, the success rates for finding an RR star, may be underestimated.
On the other hand, especially at fainter magnitudes, some stars which should have been
detectable, will not have been registered, thus overestimating the success rate.  
%For stars brighter than magnitude 14, these two effects may be assumed to compensate each other.

With the above restrictions imposed, 582 {\it NSVS} objects were identifed as GCVS RR Lyrae stars 
by their {\sc HTM} identification \citep[see][]{skydot}.
{\it NSVS} synonyms, the same object observed in overlapping 
{\it NSVS} fields, have been counted separately here.  
This will be done also for the remainder of this section, as it will not change the statistics very much.

The further restriction that there needed to be at least 100 good data points limited the sample
to 314 objects (273 RRab and 41 RRc), or 54\% of the total number of stars identified.  
Compare this to the overall 42\% of objects with at least 100 good points (8393519 out of a total of 19995106 {\it NSVS} objects).
60\% of the GCVS stars that are on average brighter than magnitude 14, have more than 100 observations,
and 68\% if only objects North of the equator are counted.

\subsection[]{Skewness}

\begin{figure}
\includegraphics[width=84mm]{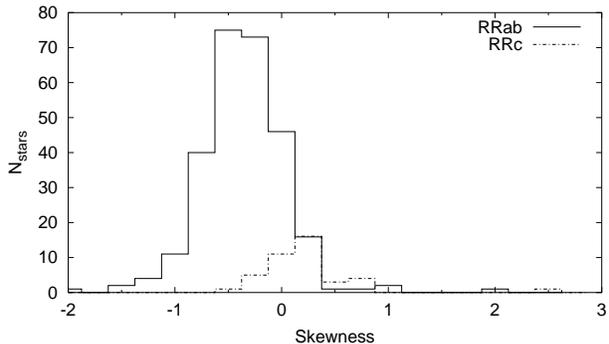}
 \caption{Distribution of the skewness values of the {\it NSVS} data for RRab and RRc stars from the GCVS.}
 \label{skew}
\end{figure}

RRab stars have a typical light curve, spending more time near minimum than near maximum, while RRc stars
have more symmetric light curves.
%This is especially the case for the RRab stars.  
It distinguishes them from eclipsing binaries, by far the most common type of variable star found in the {\it NSVS} database.
As a result, the distribution of magnitudes of an RRab star shows a negative skewness, while those of an RRc star 
shows a skewness near zero
(note that a perfect sine curve has a skewness equal to 0).  
The distribution of the skewness values for the selected GCVS stars is given in Fig.~\ref{skew}.

The maximum value for the skewness statistic was set to 0.5, which selects 303 out of 314 stars (96\%). 
Most of the GCVS objects with higher skewness values did not satisfy the standard deviation criterion either.
Therefore the chosen restriction seems to be an appropriate one.
The objects not selected include HM~Aql (probably a constant star, see below) and UU~Cam (often referred to as an EW type variable)
and a few objects near the magnitude limit resulting in a disproportionally high
number of bright observations, and hence a larger skewness.  
One star had two very faint data points, outside its normal range, most probably data errors, also influencing skewness to rise abnormally.

Some of the objects selected with positive skewness are eclipsing variables of the W~UMa type (EW).
Especially for skewness values $> 0.5$, most of the variables are EW stars or other eclipsing binaries.
Some of these have been reported by \cite{ecl}.
In a number of cases it is impossible to distinguish between EW and RRc on the basis of the light curve alone.  
These stars have not been withheld.

\subsection[]{Mean square of successive differences}

The mean square of successive differences ({\sc MSSD}) gives an indication for the timescale of the variation
of a star as compared to the sampling timescale.  
The statistic considered further in this paper is in fact the rescaled value
\begin{equation}
  \theta = 1-{\sc MSSD}/2\sigma^2,
\end{equation}
with $\sigma$ the standard deviation.  
For a purely random distribution of the data $\theta = 0$ \citep{cuypers}, 
and for a star that only brightens or fades during the total time interval, $\theta$ will approach 1.
RR Lyrae stars are rapidly changing stars (their periods range between about 0.2 and 1 day), 
compared to the frequency of observation by {\it ROTSE} (up to 5 points per night).
Therefore a raw light curve will show almost random variation.
The criterion $\theta < 0.65$ was chosen.  307 (98\%) of the selected GCVS stars
satisfy this criterion.

Because of the {\it ROTSE} observing regime, in some cases it is possible that a star
with a period close to an integer fraction of a day, will show a raw light curve that
resembles one of a much longer period star.  
This extreme aliasing is the case for RU~Boo (period 0.4927d), V1949~Cyg (0.4989d), AW~Lyr (0.4975d),
BT~Leo (0.4997d) and FY~Aqr.  For the latter, a period has not been given in the GCVS, 
but from {\it ASAS3} \citep{asasI} and {\it NSVS} data a period of 1.0229 days may be derived; 
see also \cite{dom04} for another period determination of this star.

\subsection[]{Kurtosis}

Kurtosis is a measure of the "peakedness" of a distribution and it was found
to be useful here as a discriminator against stars with extreme values.
A maximum value of 4.5 was set for the kurtosis value of the magnitudes (a perfect sine has kurtosis = 1.5). 
It excludes stars which have some unusually faint or bright data points.  
This criterion effectively removes stars with bad data points making the standard deviation erroneously large.
It may however hide true variation in some particular cases, 
as some stars with highly deviating points which are really variable are excluded as well.
This did not affect the majority of the RR Lyrae stars as the criterion selects 301 (96\%) of the GCVS stars,
All of the objects that failed this test also violated other criteria as well.
Overall with all criteria so far applied
(without limits on the standard deviation) 287 (255 RRab and 32 RRc) out of 314 (91\%) are left.

%\subsection[]{Colours}
%
%RR Lyrae stars are fairly blue stars (spectral types A and F).  
%The {\it NSVS} survey however observed only in one colour (unfiltered CCD), so colour information has to be retrieved
%from another source, such as {\it 2MASS} \citep{2mass}.  
%The {\it NSVS} positions are not very accurate, and
%matching them to {\it 2MASS} coordinates may be troublesome in crowded fields.  
%In view of this and of reddening aspects, it was decided not to use colour information as a filter.

\subsection[]{Standard deviation}

Being the most important parameter for the recognition of a variable star, 
a cut is difficult to define for the standard deviation, as it strongly depends on the average magnitude
of the stars, especially for fainter objects.  
From a Fourier fit to 
the observations, "theoretical" standard deviations (i.e. without observational errors)
for the selected RR Lyrae stars were found to be always larger than 0.1 mag, 
unless the star was in a close pair which could not be resolved by the instrument.  
Most RRab stars have "theoretical" standard deviations between 0.15 and 0.30 mag,
RRc variables between 0.10 and 0.20 mag (note that in the case of a perfect sine curve,
the standard deviation $\sigma = \Delta m/2\sqrt2$, with $\Delta m$ the total amplitude from minimum to maximum).

A survey for Cepheids in Milky Way fields of the {\it NSVS} \citep{cep}
has shown that the standard deviation needed to be at least about twice that of the average
standard deviation at the star's magnitude, to be certain the star is a genuine variable star, and not one
with unusually high observational scatter.  Therefore this restriction was chosen.
Without imposing it, the number of stars to be checked grows exponentially with decreasing standard deviation.
True variables exist which have a lower standard deviation, as shown by some of the GCVS stars,
but these cases are rather rare and/or hard to confirm, as true variation gets masked by observational scatter. 
In addition, inaccuracies on the other statistics calculated, increase as well.
 
\begin{figure}
\includegraphics[width=84mm]{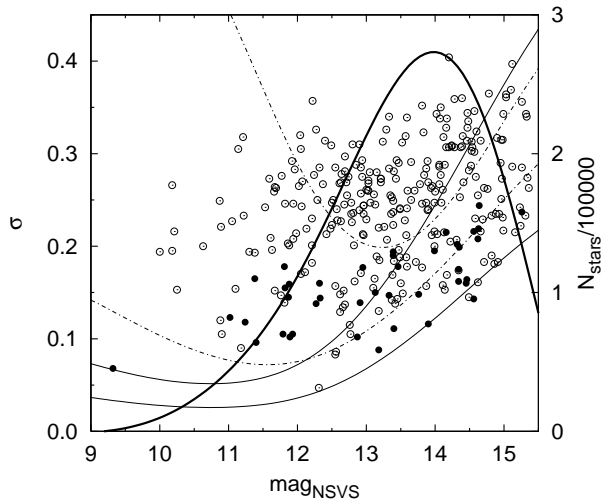}
 \caption{{\it NSVS} standard deviation compared to magnitude for the RRab (open circles) and RRc stars (filled circles) 
selected from the GCVS.
The lower and upper dashed lines represent respectively the 99 and 99.9 percentiles of the {\it NSVS} standard deviations $\sigma$ as a function of magnitude.
The bottom thin full line gives the average standard deviation for a given magnitude ($1 \sigma$ level).  
The upper thin full line is the $2 \sigma$ level and it represents the chosen cutoff limit.
The total number of stars with at least 100 data points, is given per 0.1 mag bin by the thick line (right axis).}
% \caption{{\it NSVS} standard deviation compared to magnitude for the RRab (open circles) and RRc stars (filled circles) selected from the GCVS.
%99\% of the stars have a standard deviation $\sigma$ below the dashed line, and 99.9\% below the dashed-dotted line.
%The dotted line gives the average standard deviation for a given magnitude.  
%The thin full line represents the chosen cutoff limit.
%The total number of stars with at least 100 data points, is given per 0.1 mag bin by the thick line (right axis).}
 \label{stdev}
\end{figure}

%From Fig.~\ref{stdev} it can be seen that almost all the RR Lyrae stars brighter than magnitude 13
%belong to the top 1\% of the stars with the highest standard deviation.
The cutoff for the standard deviation $\sigma$ is graphically illustrated in Fig.~\ref{stdev}.
It plots $\sigma$ for the 314 GCVS RR Lyrae stars against their average magnitude.  
The thin full lines represent the average $\sigma$ (lower line) 
and the chosen lower limit $2\sigma$ (upper line) at the given magnitude for all stars with more than 100 data points in the {\it NSVS} database.  
0.1\% of the stars have a value of $\sigma$ above the upper dashed line,
and 1\% above the lower dashed line.  Almost all the RR Lyrae stars belong to the latter group.

It is probably worth looking at the brightest stars with low standard deviation.
HM~Aql \citep{harwood} looks constant in {\it NSVS} data as well as in {\it ASAS3} data \citep{asasV}.
Also V1510~Cyg and HU~Cam appear to be constant in {\it NSVS} data.  
There might be an identification problem for these stars or a bright close companion.

\subsection[]{Detection probability}

Only 205 RRab and 16 RRc stars out of the total of 314 GCVS objects are left after applying the criterion on the standard deviation.
So the overall detection possibility for RRab stars is 75\%, for RRc stars it is only 40\%.  
However, for stars on average brighter than mag 13, the success rate is 91 and 83\% respectively.
For stars brighter than magnitude 14, these rates are reduced to 83 and 57\%.
If these numbers are then applied to the requirement that the stars needed to be observed
at least 100 times, it follows that 56\% of the RRab stars on average brighter than magnitude 14
in the northern hemisphere are detectable with the assumed criteria and nearly 40\% of the RRc stars.
These detection probabilities will be compared with the results from the {\it ASAS3} survey.
Because of the different detection possibilities, it is hard to accurately determine the
ratio between the number of RRab and RRc stars.

\section{Results}

More than 3000 objects satisfy all the criteria.  These are not all RR Lyrae stars however, some are not even genuine variables.
The final selection was based on the period and the shape of the phased light curve.
Periods were determined using the PDM technique \citep{pdm}.  For RRab stars the
default setting $N_b=5$ gives only approximate values, because the rise to maximum is so steep
both minimum and maximum observations may appear in the same bin, with an excessively large
value for the PDM statistic as a consequence.  Therefore $N_b=10$ was used. 
The RR Lyrae subtype was determined by inspection of the phased light curve: 
stars for which the ascending branch takes less than 30\% of the period were considered to be of type RRab (in most cases it is much less), others of type RRc.
Stars with a symmetric light curve, which could also be contact binaries, were rejected from the final catalogue.
All in all 785 RR Lyrae stars were detected, for which the following details
are given in the main catalogue (Table~\ref{maintable}; the complete table is available electronically only): {\it NSVS} number, 
coordinates in degrees, magnitude range (brightest and faintest observed magnitude), type, period in days, epoch of maximum light (HJD - 2450000) 
and a cross identification if one exists.
References for some of the abbreviations used in the cross identifications are given below the table.  
{\it NSVS} synonyms have not been included in this table.
  
714 of these stars are of type RRab (of which 469 are brighter than magnitude 14 on average),
65 of type RRc (all brighter than magnitude 14; not surprisingly since the lower limit for the standard deviation
at magnitude 14 is larger than the average standard deviation for RRc stars) and 6 are of type RRd.  
Of the 785 stars, 188 are previously unknown.  
Many others have only been suspected of variability before or had been wrongly classified.  
For others, it is the first time a period has been given.
The catalogue contains 342 RR Lyrae stars already known in the GCVS and 23 stars with an incorrect type in the GCVS.
However it does not contain 178 GCVS RR Lyrae stars with 50 or more good data points that could be confirmed to be RR Lyrae stars
from the {\it NSVS} data, but that did not satisfy all of the selection criteria given above.
For completeness these variables are given in Table~\ref{missed} (available in full electronically only), with the same layout as Table~\ref{maintable}.

\subsection{Blazhko stars}

Some RR Lyrae stars show a cyclic modulation of the amplitude and shape of their light curves
known as the Blazhko effect \citep{blazhko}.  
The effect manifests itself as one or two additional frequencies in the periodogram, very close to the main frequency.
The total timespan of available {\it NSVS} data is only about 9 months, which is short
to accurately determine Blazhko periods.  However, it can provide reasonable indications, as a number
of the known Blazhko stars were easily picked up, and gave Blazhko periods in good agreement
with the values from the literature.  
All stars from Tables~\ref{maintable} and \ref{missed} were checked for a Blazhko effect.
Objects for which no effect was found include 
SW~And, RS~Boo (known Blazhko period $>500$ days, so it is not surprising its effect was not noticed), 
SW~Boo, XZ and DM~Cyg, XZ~Dra, RU~Psc, and RR~Lyr itself (in the latter case too few data points are available).
It has been impossible to detect a Blazhko effect in stars fainter than about magnitude 13.5.

The results are summarized in Table~\ref{blazhko}.  Stars for which at least one linear combination of the main frequency and
the additional frequency have been found, can be positively identified as Blazhko stars.  
Stars for which no such combination has been detected are indicated with a ":" after the Blazhko period, and need further confirmation.
The stars which are known to be Blazhko stars in the GCVS are marked with an asterisk after their name.  
Except for RZ~Lyr and AR~Ser, the Blazhko periods given are not much different from those derived here.
For RZ~Lyr the Blazhko period given in the GCVS is 59 days.  
For AR~Ser, a variable Blazhko period between 80 and 120 days has been given.
\cite{lee1} give a Blazhko period of 122 days for FM~Per and 57.5 days for DR~And \citep{lee2}.
\cite{lee0} already noted the Blazhko effect in V421~Her.
Finally, the Blazhko effect of OV~And has been studied in more detail by \cite{ovand}, but the data were inconclusive.

\subsection{Double mode stars}

Galactic field double-mode RR Lyrae stars (RRd) are very rare with less than twenty currently
known. These are low-amplitude variables very similar to RRc
stars but showing two periods, the fundamental $P_0$ and first overtone,
$P_1$, with a ratio $P_1/P_0 \approx 0.744$.  In most cases, the first overtone mode has the highest amplitude.
Six RRd stars have been detected.
The double-mode nature of V372~Ser \citep{enrique} was already known before.  The true nature of
GSC~3047-0176 \citep{koppelman}, GSC~3059-0636 \citep{smith} and GSC~4868-0831 \citep{rrdasas} was recently found in other studies
of the {\it NSVS} or {\it ASAS3} data.  The two remainig RRd stars are identified here for the first time.
Table~\ref{rrd} gives the fundamental period $P_0$ and the first overtone period $P_1$, 
as well as the period ratio for the six double mode stars.

\subsection{Misclassified GCVS stars}\label{misclass}

Ten stars classified as RR Lyrae variables in the GCVS, turned
out to be of another type.  
V1180~Aql is a close double \citep{hoffmeister}, one star of the pair is a Mira variable, not an RR.
AU and V556~Cas, V811~Oph, V421~Per and BQ~Pup are Cepheids and 
V1823~Cyg and IT~Her are probably EW-type stars.
V1069~Sgr may be constant or its position may be in error.
Also BU~UMa is most likely a constant star (P. Van Cauteren, private communication).

23 genuine RR Lyrae stars (included in Table~\ref{maintable}) are also misclassified in the GCVS.  
These include DP~Aqr, AL and TZ~Cap,
V939~Cyg, SX and FM~Del, AQ and KO~Dra, V534~Her and V1429~Oph.  
Others like V344~Ser and KQ~UMa have been identified earlier also by \cite{khruslov}. 

%There are a few errors in the RR Lyrae subtype as well, e.g. EZ~Cep, classified as RRc, is in fact an RRab.

\subsection{Period corrections}

Some stars in the GCVS are correctly classified but have a wrong period, mostly an alias of the correct period.
These stars include GT~Aqr \citep{diethelm}, X~CMi, RW~Equ, DG~Hya, V418~Her and V784~Oph. 

\subsection{Estimates of the number of RR Lyrae stars}

From the analysis and the detection possibilities discussed above, 
it is found that there should be about 650 RRab stars in the northern hemisphere brighter than magnitude 14
(of which 365 have been detected in this study) and about 140 RRc stars (56 have been detected).
One could state that about four out of five galactic field RR Lyrae stars are of type RRab.
However, because of selection effects which favour RRab stars, 
the exact fraction of the different subtypes are uncertain.  
Only about 2\% of the RR Lyrae stars are double-mode stars.  
But because of the low number of RRd stars, this frequency is even more uncertain.

These estimates can be compared to those obtained from a comparison of the current catalogue 
with the results of the All Sky Automated Survey \citep[{\it ASAS3}, ][]{asasV}.  
That survey is based in the southern sky and detected variables south of declination $\sim +28$.  
In the overlapping region from declination $-15$ to $+15$, {\it ASAS3} detected 323 objects unambiguously identified as RRab stars on average brighter than magnitude 14 
(assuming that the average magnitude of an RRab star equals the maximum magnitude plus 0.65 times the full amplitude),
and 73 RRab stars fainter than magnitude 14.  Of these respectively 150 and 28 were detected in the {\it NSVS} data, giving
detection probabilities of 46 and 38\%.  
These probabilities do not change when only the region from declination $-10$ to $+10$ is considered.
However if the declination zone between $0$ and $+15$ is considered only, the detection probability for RRab stars brighter
than magnitude 14 rises to 53\% (89 out of 168 stars were identified), very close to the 56\% found from the GCVS sample.
It is clear that the efficiency of the search deteriorates fast for negative declinations, 
probably for a large part due to the requirement of at least 100 data points.

The estimate for the RRc stars differs more.
{\it ASAS3} identified 83 stars unambiguously as RRc stars in the considered overlapping region (all brighter than magnitude 14), 
of which only 10 were found in the {\it NSVS} data.
This results in a detection probability of only 12\% (14\% if the region from declination $-10$ to $+10$ is considered,
and 19\% for stars in the declination zone between $0$ and $+15$). 
The large difference with the estimate from the GCVS sample can probably be explained by the small number of RRc stars found overall, 
so that the chosen samples are not really representative for the total population.
Again, it can be concluded that the estimate for the number of RRc stars is not very reliable, 
and may in reality be considerably higher.

Figures for the southern sky can be obtained from the reverse comparison.  Of the 124 RRab stars brighter than magnitude 14
between declination $-15$ to $+15$ found in this study, 114 were identified as well by {\it ASAS3} 
(including four that have an ambiguous variability type,
but not including five stars with a different type in {\it ASAS3}).  This results in a detection probability of 92\% for {\it ASAS3}.  
As the survey unambiguously identified 817 objects south of the equator and brighter than magnitude 14 as RRab stars (1015 if ambiguous types are considered as well), 
there will be about 890 in total (or 1100 when all the ambiguously typed stars are taken into account).
It may be concluded that there is an over-abundance of RRab stars in the southern sky compared to the northern sky.

For the RRc stars the figures are as follows: of the 16 RRc stars found in this study in the overlapping region
between  declinations $-15$ and $+15$, 
12 or 75\% were identified by {\it ASAS3} as RR Lyrae stars as well.
With 259 stars brighter than magnitude 14 unambiguously identified as RRc stars,
there may be about 345 in total south of the equator.  
This may be considered as a lower limit as there are 1047 possible RRc stars given by {\it ASAS3}. 
However a large part of those are probably not genuine RRc stars.

\section{Conclusion}

This paper has enlarged the known number of galactic field RR Lyrae stars in the northern sky.  
The number of known stars showing multiperiodicity (Blazhko effect and double-mode
pulsation) has been enlarged as well.  The number of RR Lyrae stars brighter than magnitude 14 has been estimated,
showing that there is an over-abundance of RRab stars in the southern hemisphere.

\section*{Acknowledgments}
The authors thank John Greaves, Sebasti\'an Otero, Gisela Maintz and Christoph Kaser for helpful suggestions.
Doug Welch is acknowledged for granting access to the copy of the {\it NSVS} database at McMaster University.
This publication makes use of the data from the Northern Sky Variability Survey 
created jointly by the Los Alamos National Laboratory and University of Michigan.
The {\it NSVS} was funded by the US Department of Energy, 
the National Aeronautics and Space Administration and the National Science Foundation.
This study also used the SIMBAD and VizieR 
databases operated at the {\it Centre de Donn\'ees Astronomiques (Strasbourg)} in France.

\clearpage

\setcounter{table}{0}
\begin{table}
 \caption{RR Lyrae objects identified in the NSVS database}
 \label{maintable}
\begin{tabular}{rrrrlrcl}
  \hline
NSVS ID & \multicolumn{2}{c}{RA (2000) Dec} & NSVS Mag. & Type & \multicolumn{1}{c}{Period} & Epoch & Cross ID \\ 
        & \multicolumn{2}{c}{[degrees]} & & & \multicolumn{1}{c}{[days]} & \multicolumn{1}{c}{HJD-2450000} & \\
  \hline
 6313547 &   0.0164 & +35.3631 & 12.71-13.88 & RRab & 0.70669 & 1465.50 & GM And \\ 
14654207 &   0.9480 & -11.4764 & 12.58-13.62 & RRab & 0.74084 & 1438.52 & ASAS000348-1128.6 \\ 
17443023 &   1.5857 & -35.2884 & 13.43-14.12 & RRab & 0.36595 & 1487.71 & NSV 29 \\ 
 1616266 &   3.8114 & +60.3403 & 10.38-10.77 & RRc  & 0.54647 & 1505.61 & V363 Cas \\ 
 9088466 &   4.4523 & + 9.8886 & 11.94-12.56 & RRab & 0.70119 & 1462.67 & ASAS001748+0953.3 \\ 
 3656843 &   5.1850 & +40.8276 & 10.90-11.26 & RRab & 0.47060 & 1403.69 & OV And \\
\hline
 \end{tabular}
\end{table}

\clearpage

\setcounter{table}{0}
\begin{table}
 \caption{continued}
 \begin{tabular}{rrrrlrcl}
  \hline
NSVS ID & \multicolumn{2}{c}{RA (2000) Dec} & NSVS Mag. & Type & \multicolumn{1}{c}{Period} & Epoch & Cross ID \\ 
        & \multicolumn{2}{c}{[degrees]} & & & \multicolumn{1}{c}{[days]} & \multicolumn{1}{c}{HJD-2450000} & \\
  \hline
 9253268 &  32.2680 & +21.1914 & 13.83-14.97 & RRab & 0.47157 & 1467.68 & TU Ari \\ 
 3939460 &  33.6697 & +49.8887 & 12.48-13.01 & RRab & 0.55289 & 1425.75 &  \\ 
14762124 &  33.8121 & -10.8001 & 10.70-11.36 & RRab & 0.62348 & 1521.73 & RV Cet \\ 
 9240848 &  34.9677 & + 9.2793 & 12.77-13.39 & RRab & 0.46273 & 1454.83 & ASAS021952+0916.8 \\ 
14768131 &  37.1354 & - 8.3582 & 11.51-12.42 & RRab & 0.51066 & 1490.19 & RZ Cet \\ 
 3994861 &  37.6315 & +40.8429 & 13.17-14.28 & RRab & 0.60687 & 1558.66 & DU And \\
\hline
 \end{tabular}
\end{table}

\clearpage

\setcounter{table}{0}
\begin{table}
 \caption{continued}
 \begin{tabular}{rrrrlrcl}
  \hline
NSVS ID & \multicolumn{2}{c}{RA (2000) Dec} & NSVS Mag. & Type & \multicolumn{1}{c}{Period} & Epoch & Cross ID \\ 
        & \multicolumn{2}{c}{[degrees]} & & & \multicolumn{1}{c}{[days]} & \multicolumn{1}{c}{HJD-2450000} & \\
  \hline
 9753568 &  95.6834 & +18.5305 & 12.27-12.71 & RRab & 0.28671 & 1540.80 & ASAS062244+1831.9 \\ 
 2336392 &  98.4985 & +67.0255 & 12.47-13.73 & RRab & 0.48042 & 1541.73 & RZ Cam \\ 
15206917 &  98.6774 & -15.7776 & 13.19-14.54 & RRab & 0.40858 & 1486.87 & ASAS063442-1546.7 \\ 
15172858 & 101.3309 & - 8.8388 & 12.77-13.64 & RRab & 0.41343 & 1563.76 & DV Mon \\ 
 4694846 & 107.8960 & +40.7769 & 11.47-12.73 & RRab & 0.39166 & 1593.62 & TZ Aur \\ 
 9934090 & 108.0236 & +17.3801 & 13.69-14.37 & RRab & 0.72400 & 1504.71 & NSV 3449 \\
\hline
 \end{tabular}
\end{table}

\clearpage

\setcounter{table}{0}
\begin{table}
 \caption{continued}
 \begin{tabular}{rrrrlrcl}
  \hline
NSVS ID & \multicolumn{2}{c}{RA (2000) Dec} & NSVS Mag. & Type & \multicolumn{1}{c}{Period} & Epoch & Cross ID \\ 
        & \multicolumn{2}{c}{[degrees]} & & & \multicolumn{1}{c}{[days]} & \multicolumn{1}{c}{HJD-2450000} & \\
  \hline
 2496194 & 139.3336 & +68.6355 & 14.18-15.34 & RRab & 0.48599 & 1275.81 & KQ UMa \\ 
 7472263 & 139.7749 & +29.0656 & 11.40-12.48 & RRab & 0.54720 & 1274.73 & RW Cnc \\ 
 7454553 & 139.9645 & +33.8733 & 13.50-14.24 & RRab & 0.52174 & 1577.67 &  \\ 
15679507 & 140.5880 & -13.6473 & 13.07-14.10 & RRab & 0.53218 & 1502.92 & IV Hya \\ 
 4838955 & 141.1273 & +42.3047 & 12.79-13.51 & RRab & 0.56211 & 1606.66 &  \\ 
15681946 & 141.2336 & -13.2002 & 13.27-14.46 & RRab & 0.61129 & 1558.78 & ASAS092456-1312.0 \\
\hline
 \end{tabular}
\end{table}

\clearpage

\setcounter{table}{0}
\begin{table}
 \caption{continued}
 \begin{tabular}{rrrrlrcl}
  \hline
NSVS ID & \multicolumn{2}{c}{RA (2000) Dec} & NSVS Mag. & Type & \multicolumn{1}{c}{Period} & Epoch & Cross ID \\ 
        & \multicolumn{2}{c}{[degrees]} & & & \multicolumn{1}{c}{[days]} & \multicolumn{1}{c}{HJD-2450000} & \\
  \hline
10356449 & 172.2964 & + 9.5362 & 13.46-15.01 & RRab & 0.54505 & 1288.89 & AS Leo \\ 
15901811 & 172.4048 & -15.7028 & 13.33-14.18 & RRab & 0.66435 & 1572.91 & ASAS112937-1542.2 \\ 
10357384 & 172.7233 & +13.3245 & 13.05-13.98 & RRab & 0.33241 & 1573.86 & ASAS113053+1319.4 \\ 
13138797 & 173.0785 & + 3.2874 & 13.94-14.79 & RRab & 0.51509 & 1582.87 & ASAS113219+0317.2 \\ 
10358222 & 173.2656 & +12.1538 & 12.13-12.89 & RRab & 0.72683 & 1332.71 & AX Leo \\ 
 4974265 & 173.9029 & +38.7660 & 11.83-12.27 & RRc  & 0.26796 & 1633.68 & GSC 3013-01988 $^3$ \\
\hline
 \end{tabular}
\end{table}

\clearpage

\setcounter{table}{0}
\begin{table}
 \caption{continued}
 \begin{tabular}{rrrrlrcl}
  \hline
NSVS ID & \multicolumn{2}{c}{RA (2000) Dec} & NSVS Mag. & Type & \multicolumn{1}{c}{Period} & Epoch & Cross ID \\ 
        & \multicolumn{2}{c}{[degrees]} & & & \multicolumn{1}{c}{[days]} & \multicolumn{1}{c}{HJD-2450000} & \\
  \hline
 7651881 & 195.3717 & +32.0869 & 13.77-15.15 & RRab & 0.55187 & 1618.39 & TZ CVn \\ 
 7666039 & 195.5569 & +24.2388 & 13.17-14.42 & RRab & 0.66141 & 1318.70 & BF Com \\ 
  960209 & 195.8796 & +71.1122 & 10.40-10.73 & RRc  & 0.33297 & 1280.64 &  \\ 
10447559 & 197.0760 & +18.5406 & 13.45-14.96 & RRab & 0.54666 & 1616.26 & Z Com \\ 
 2690430 & 197.1737 & +65.6668 & 12.39-12.80 & RRc  & 0.31299 & 1390.73 &  \\ 
10433141 & 197.8103 & +11.1578 & 13.66-14.48 & RRab & 0.53627 & 1307.75 & ASAS131114+1109.5 \\
\hline
 \end{tabular}
\end{table}

\clearpage

\setcounter{table}{0}
\begin{table}
 \caption{continued}
 \begin{tabular}{rrrrlrcl}
  \hline
NSVS ID & \multicolumn{2}{c}{RA (2000) Dec} & NSVS Mag. & Type & \multicolumn{1}{c}{Period} & Epoch & Cross ID \\ 
        & \multicolumn{2}{c}{[degrees]} & & & \multicolumn{1}{c}{[days]} & \multicolumn{1}{c}{HJD-2450000} & \\
  \hline
13343222 & 216.8725 & + 3.7795 & 12.90-14.09 & RRab & 0.63387 & 1607.89 & AE Vir \\ 
 7752385 & 216.8958 & +36.0457 & 12.09-13.16 & RRab & 0.51354 & 1616.22 & SW Boo \\ 
13322485 & 216.9128 & - 0.9014 & 11.23-12.32 & RRab & 0.41082 & 1341.19 & ST Vir \\ 
13343644 & 217.0412 & + 6.5455 & 11.38-12.35 & RRab & 0.48372 & 1311.78 & AF Vir \\ 
 5129241 & 217.6285 & +50.3393 & 13.17-13.85 & RRab & 0.57067 & 1567.83 &  \\ 
10526447 & 217.7533 & +20.4249 & 13.71-14.48 & RRab & 0.61712 & 1631.76 &  \\
\hline
 \end{tabular}
\end{table}

\clearpage

\setcounter{table}{0}
\begin{table}
 \caption{continued}
 \begin{tabular}{rrrrlrcl}
  \hline
NSVS ID & \multicolumn{2}{c}{RA (2000) Dec} & NSVS Mag. & Type & \multicolumn{1}{c}{Period} & Epoch & Cross ID \\ 
        & \multicolumn{2}{c}{[degrees]} & & & \multicolumn{1}{c}{[days]} & \multicolumn{1}{c}{HJD-2450000} & \\
  \hline
 7818286 & 229.2717 & +35.1157 & 11.92-13.10 & RRab & 0.45694 & 1286.72 & UU Boo \\ 
13399252 & 229.3958 & - 1.0889 & 11.18-11.88 & RRd  & 0.35072 & 1604.00 & V372 Ser \\ 
 5181225 & 229.5150 & +46.7040 & 11.14-11.69 & RRab & 0.45866 & 1275.69 & DG Boo \\ 
13399707 & 229.5561 & - 1.3191 & 13.68-14.46 & RRab & 0.60488 & 1617.97 & FASTT 701 \\ 
16237482 & 229.5912 & - 8.4619 & 11.61-12.67 & RRab & 0.26963 & 1614.00 & TV Lib \\ 
13425078 & 229.8271 & + 7.8869 & 12.65-13.48 & RRab & 0.54764 & 1337.79 & ASAS151919+0753.1 \\
\hline
 \end{tabular}
\end{table}

\clearpage

\setcounter{table}{0}
\begin{table}
 \caption{continued}
 \begin{tabular}{rrrrlrcl}
  \hline
NSVS ID & \multicolumn{2}{c}{RA (2000) Dec} & NSVS Mag. & Type & \multicolumn{1}{c}{Period} & Epoch & Cross ID \\ 
        & \multicolumn{2}{c}{[degrees]} & & & \multicolumn{1}{c}{[days]} & \multicolumn{1}{c}{HJD-2450000} & \\
  \hline
 5237136 & 244.8092 & +39.5020 & 13.30-14.75 & RRab & 0.48014 & 1451.12 & WX CrB \\ 
 7879638 & 244.8578 & +29.7133 & 11.48-12.02 & RRc  & 0.33162 & 1278.72 & RV CrB \\ 
10664602 & 245.1852 & + 9.7402 & 14.37-15.47 & RRab & 0.47708 & 1356.77 &  \\ 
 2824910 & 245.2730 & +58.4503 & 12.53-13.32 & RRc  & 0.32106 & 1414.68 & VZ Dra \\ 
 2869677 & 246.4965 & +65.2371 & 14.05-15.06 & RRab & 0.62220 & 1328.73 &  \\ 
13544287 & 247.0452 & + 3.0701 & 13.45-14.48 & RRab & 0.59701 & 1312.77 & ASAS162811+0304.3 \\
\hline
 \end{tabular}
\end{table}

\clearpage

\setcounter{table}{0}
\begin{table}
 \caption{continued}
 \begin{tabular}{rrrrlrcl}
  \hline
NSVS ID & \multicolumn{2}{c}{RA (2000) Dec} & NSVS Mag. & Type & \multicolumn{1}{c}{Period} & Epoch & Cross ID \\ 
        & \multicolumn{2}{c}{[degrees]} & & & \multicolumn{1}{c}{[days]} & \multicolumn{1}{c}{HJD-2450000} & \\
  \hline
10824662 & 260.6612 & +17.8844 & 12.21-13.68 & RRab & 0.43605 & 1600.42 & V394 Her \\ 
10824626 & 260.7845 & +22.6587 & 13.24-14.59 & RRab & 0.55620 & 1325.77 & V397 Her \\ 
13688631 & 260.8057 & + 1.8644 & 12.20-12.56 & RRab & 0.73777 & 1352.79 & ASAS172313+0151.8 \\ 
13644721 & 260.9004 & - 1.1914 & 12.14-12.88 & RRab & 0.66793 & 1403.76 & ASAS172336-0111.5 \\ 
 7962128 & 260.9231 & +37.3258 & 13.10-13.96 & RRab & 0.45402 & 1478.65 &  \\ 
 8038923 & 262.1327 & +38.3771 & 14.34-15.30 & RRab & 0.45834 & 1415.83 & ROTSE1\_J172831.79+382236.9 \\
\hline
 \end{tabular}
\end{table}

\clearpage

\setcounter{table}{0}
\begin{table}
 \caption{continued}
 \begin{tabular}{rrrrlrcl}
  \hline
NSVS ID & \multicolumn{2}{c}{RA (2000) Dec} & NSVS Mag. & Type & \multicolumn{1}{c}{Period} & Epoch & Cross ID \\ 
        & \multicolumn{2}{c}{[degrees]} & & & \multicolumn{1}{c}{[days]} & \multicolumn{1}{c}{HJD-2450000} & \\
  \hline
 3038588 & 279.4839 & +56.8290 & 13.62-14.54 & RRab & 0.59435 & 1475.61 &  \\ 
 3040364 & 279.9083 & +58.1000 & 13.95-14.87 & RRab & 0.52746 & 1389.73 &  \\ 
 5494746 & 280.0971 & +41.0403 & 13.06-14.12 & RRab & 0.54555 & 1603.96 & LX Lyr \\ 
 8178485 & 280.3172 & +28.7233 & 11.23-11.61 & RRab & 0.41141 & 1360.13 & CN Lyr \\ 
 8234128 & 280.9082 & +32.7979 & 11.20-12.35 & RRab & 0.51131 & 1600.99 & RZ Lyr \\ 
11061929 & 281.2414 & +23.0543 & 12.61-13.85 & RRab & 0.53722 & 1307.44 & V347 Her \\
\hline
 \end{tabular}
\end{table}

\clearpage

\setcounter{table}{0}
\begin{table}
 \caption{continued}
 \begin{tabular}{rrrrlrcl}
  \hline
NSVS ID & \multicolumn{2}{c}{RA (2000) Dec} & NSVS Mag. & Type & \multicolumn{1}{c}{Period} & Epoch & Cross ID \\ 
        & \multicolumn{2}{c}{[degrees]} & & & \multicolumn{1}{c}{[days]} & \multicolumn{1}{c}{HJD-2450000} & \\
  \hline
 8588795 & 312.4413 & +24.2144 & 11.83-12.25 & RRc  & 0.33609 & 1324.89 & ASAS204946+2412.8 \\ 
14311711 & 312.9558 & + 2.3913 & 13.94-15.40 & RRab & 0.44490 & 1450.63 & DP Aqr \\ 
11491144 & 313.0251 & +21.8229 & 13.38-13.95 & RRc  & 0.31332 & 1337.78 & ASAS205206+2149.3 \\ 
11491797 & 313.1282 & +22.4361 & 12.23-13.08 & RRab & 0.43406 & 1496.13 & FK Vul \\ 
14352725 & 313.2229 & - 0.6331 & 13.10-13.81 & RRab & 0.33361 & 1421.77 & ASAS205254-0038.0 \\ 
 3306145 & 313.7241 & +66.4448 & 12.02-12.89 & RRab & 0.46789 & 1322.87 & FP Cep \\
\hline
 \end{tabular}
\end{table}

\clearpage

\setcounter{table}{0}
\begin{table}
 \caption{continued}
 \begin{tabular}{rrrrlrcl}
  \hline
NSVS ID & \multicolumn{2}{c}{RA (2000) Dec} & NSVS Mag. & Type & \multicolumn{1}{c}{Period} & Epoch & Cross ID \\ 
        & \multicolumn{2}{c}{[degrees]} & & & \multicolumn{1}{c}{[days]} & \multicolumn{1}{c}{HJD-2450000} & \\
  \hline
 6094125 & 335.2934 & +39.7200 & 12.13-13.30 & RRab & 0.62000 & 1438.36 & CQ Lac \\ 
11734397 & 335.5116 & + 8.8227 & 14.35-15.31 & RRab & 0.60895 & 1421.76 &  \\ 
14529130 & 335.9223 & + 4.1039 & 13.61-14.40 & RRab & 0.51799 & 1475.64 &  \\ 
14504106 & 336.2604 & - 4.9495 & 13.64-15.09 & RRab & 0.52778 & 1414.75 & FH Aqr \\ 
 8895403 & 336.3465 & +34.8858 & 13.02-14.23 & RRab & 0.50344 & 1361.78 & GY Peg \\ 
11770210 & 336.3669 & +20.1519 & 13.14-13.81 & RRab & 0.64271 & 1421.76 &  \\
\hline
 \end{tabular}
\end{table}

\clearpage

\setcounter{table}{0}
\begin{table}
 \caption{continued}
 \begin{tabular}{rrrrlrcl}
  \hline
NSVS ID & \multicolumn{2}{c}{RA (2000) Dec} & NSVS Mag. & Type & \multicolumn{1}{c}{Period} & Epoch & Cross ID \\ 
        & \multicolumn{2}{c}{[degrees]} & & & \multicolumn{1}{c}{[days]} & \multicolumn{1}{c}{HJD-2450000} & \\
  \hline
 6248008 & 357.6928 & +33.3512 & 12.87-13.73 & RRab & 0.53393 & 1402.87 &  \\ 
14624322 & 358.5058 & - 7.6773 & 12.46-13.12 & RRab & 0.52890 & 1401.88 & NSV 14744 \\ 
11895079 & 358.5340 & + 0.9635 & 10.76-11.14 & RRc  & 0.30623 & 1463.69 & NSV 14745 \\ 
 9075933 & 359.0035 & +10.8884 & 11.93-12.60 & RRab & 0.56249 & 1382.75 & ASAS235601+1053.3 \\ 
 3633664 & 359.6761 & +41.4886 & 12.94-14.15 & RRab & 0.60298 & 1521.65 & DY And \\ 
  \hline
 \end{tabular}

 \medskip
 \begin{tabular}{rp{200mm}}
\multicolumn{2}{l}{References to cross id's:}\\
$^1$ & \cite{rrdasas} \\
$^2$ & \cite{nichoejv} \\ 
$^3$ & \cite{pz} \\
$^4$ & \cite{koppelman} \\
$^5$ & \cite{smith} \\
$^6$ & \cite{nicholson} \\
$^7$ & \cite{bernasconi} \\
ASAS & \cite{asasV} \\
FASTT & \cite{fastt} \\
GALEX & \cite{galex} \\
MisV & http://www.aerith.net/misao \\
ROTSE & \cite{rotse} \\
QUEST & \cite{quest} \\
 \end{tabular}
\end{table}

\clearpage

\setcounter{table}{1}
\begin{table}
 \caption{Confirmed RR Lyrae stars from the GCVS excluded by the selection criteria}
 \label{missed}
 \begin{tabular}{rrrrlrcl}
  \hline
NSVS ID & \multicolumn{2}{c}{RA (2000) Dec} & NSVS Mag. & Type & \multicolumn{1}{c}{Period} & Epoch & Cross ID \\ 
        & \multicolumn{2}{c}{[degrees]} & & & \multicolumn{1}{c}{[days]} & \multicolumn{1}{c}{HJD-2450000} & \\
  \hline
14641662 &   1.0211 & -16.9979 & 12.03-12.54 & RRab & 0.60589 & 1456.70 & UU Cet \\ 
14682480 &   8.4095 & -15.4874 & 11.27-11.94 & RRab & 0.57347 & 1514.73 & RX Cet \\ 
17478336 &  13.9943 & -26.3831 & 13.27-14.28 & RRab & 0.52074 & 1402.95 & UV Scl \\ 
 3763277 &  15.7183 & +45.1897 & 14.28-15.66 & RRab & 0.48525 & 1492.84 & IY And \\ 
 3860939 &  19.6106 & +50.6715 & 13.03-13.61 & RRc  & 0.37355 & 1456.63 & V830 Cas \\ 
17510932 &  20.7232 & -26.2927 & 13.25-14.51 & RRab & 0.45151 & 1536.65 & RW Scl \\
\hline
 \end{tabular}
\end{table}

\clearpage

\setcounter{table}{1}
\begin{table}
 \caption{continued}
 \begin{tabular}{rrrrlrcl}
  \hline
NSVS ID & \multicolumn{2}{c}{RA (2000) Dec} & NSVS Mag. & Type & \multicolumn{1}{c}{Period} & Epoch & Cross ID \\ 
        & \multicolumn{2}{c}{[degrees]} & & & \multicolumn{1}{c}{[days]} & \multicolumn{1}{c}{HJD-2450000} & \\
  \hline
18691063 & 188.9169 & -26.6618 & 12.78-13.57 & RRab & 0.57738 & 1318.70 & EM Hya \\ 
 7643255 & 188.9835 & +37.2072 & 12.37-13.18 & RRab & 0.66815 & 1288.71 & SV CVn \\ 
10404312 & 189.0420 & +22.3952 & 14.08-14.94 & RRab & 0.64553 & 1310.90 & AO Com \\ 
 7656842 & 189.4881 & +29.9685 & 14.07-15.41 & RRab & 0.47242 & 1311.73 & FV Com \\ 
10405541 & 189.8281 & +22.0545 & 14.11-15.30 & RRab & 0.59366 & 1631.75 & AP Com \\ 
 7644920 & 190.2293 & +37.0855 & 12.44-13.69 & RRab & 0.44166 & 1307.73 & SW CVn \\
\hline
 \end{tabular}
\end{table}

\clearpage

\setcounter{table}{1}
\begin{table}
 \caption{continued}
 \begin{tabular}{rrrrlrcl}
  \hline
NSVS ID & \multicolumn{2}{c}{RA (2000) Dec} & NSVS Mag. & Type & \multicolumn{1}{c}{Period} & Epoch & Cross ID \\ 
        & \multicolumn{2}{c}{[degrees]} & & & \multicolumn{1}{c}{[days]} & \multicolumn{1}{c}{HJD-2450000} & \\
  \hline
 5329963 & 257.7005 & +41.6200 & 14.30-15.46 & RRab & 0.52792 & 1495.66 & V721 Her \\ 
 5331289 & 257.8341 & +44.9010 & 14.07-14.98 & RRab & 0.62099 & 1390.68 & V794 Her \\ 
10815361 & 258.6470 & +22.3959 & 14.10-15.26 & RRab & 0.67960 & 1325.74 & V379 Her \\ 
 5334360 & 259.4222 & +41.3285 & 14.11-15.72 & RRab & 0.57093 & 1328.74 & V727 Her \\ 
13682138 & 259.6576 & + 8.2280 & 13.27-14.19 & RRab & 0.50969 & 1373.78 & V864 Oph \\ 
 7995290 & 261.6600 & +26.9379 & 12.99-13.82 & RRab & 0.80619 & 1300.89 & V486 Her \\
\hline
 \end{tabular}
\end{table}

\clearpage

\begin{table}
 \caption{RR Lyrae stars showing the Blazhko effect}
 \label{blazhko}
 \begin{tabular}{lrrrrc}
  \hline
Name & NSVS ID & \multicolumn{2}{c}{RA (2000) Dec} & \multicolumn{1}{c}{Period} & Blazhko period  \\ 
     &  & \multicolumn{2}{c}{[degrees]} & \multicolumn{1}{c}{[days]} & \multicolumn{1}{c}{[days]} \\
  \hline
OV And        &  3656843 &   5.1850 & +40.8276 & 0.47060 & ~27: \\
GSC 0607-0591 &  9149730 &  11.9844 & +11.7048 & 0.45570 & ~55: \\
DR And        &  6429832 &  16.2945 & +34.2184 & 0.56307 & ~57: \\
UX Tri        &  6471887 &  26.3959 & +31.3801 & 0.46690 & ~45~ \\
RV Cet        & 14762124 &  33.8121 & -10.8001 & 0.62340 & 117~ \\
FM Per        &  4241456 &  60.8624 & +47.9978 & 0.48927 & ~20~ \\
\hline
 \end{tabular}
\end{table}

\clearpage

\begin{table}
 \caption{Double mode RR Lyrae stars}
 \label{rrd}
 \begin{tabular}{lrrrccc}
  \hline
Name & NSVS ID & \multicolumn{2}{c}{RA (2000) Dec} & \multicolumn{1}{c}{$P_0$} & \multicolumn{1}{c}{$P_1$} & Period ratio  \\ 
     &  & \multicolumn{2}{c}{[degrees]} & \multicolumn{1}{c}{[days]} & \multicolumn{1}{c}{[days]} & \\
  \hline
GSC 4868-0831 & 12919472 & 131.9456 & - 3.6501 & 0.56392 & 0.42081 & 0.7462 \\
GSC 3047-0176 &  5152353 & 223.4165 & +40.5286 & 0.47462 & 0.35298 & 0.7437 \\
V372 Ser      & 13399252 & 229.3958 & - 1.0889 & 0.47135 & 0.35072 & 0.7441 \\
GSC 3059-0636 &  5222076 & 236.6077 & +44.3125 & 0.49404 & 0.36691 & 0.7427 \\
%GSC 0406-0147 & 13664588 & 256.2904 & + 4.2275 & 0.55288 & 0.41092 & 0.7432 \\
V458 Her      & 10809333 & 257.1295 & +18.5213 & 0.48374 & 0.35998 & 0.7442 \\
GSC 4421-1234 &  1088957 & 258.3326 & +69.1317 & 0.54080 & 0.40320 & 0.7456 \\
  \hline
 \end{tabular}
\end{table}

\label{lastpage}

\end{document}